\documentclass[
superscriptaddress,
nofootinbib,
noeprint,
nobibnotes,
amsmath,amssymb,
aps,
prb,
twocolumn,
floatfix,
]{revtex4-2}

\usepackage{graphicx}
\usepackage{dcolumn}
\usepackage{bm}

\usepackage{epstopdf}
\usepackage{braket}
\usepackage{xcolor}

\begin{document}

\title{Experimental speedup of quantum dynamics through squeezing}

\author{S. C. Burd}
\thanks{These authors contributed equally to this work.}
\affiliation{Time and Frequency Division, National Institute of Standards and Technology, 325 Broadway, Boulder, Colorado 80305, USA}
\affiliation{Department of Physics, University of Colorado, Boulder, Colorado 80309, USA}

\author{H. M. Knaack}
\thanks{These authors contributed equally to this work.}
\affiliation{Time and Frequency Division, National Institute of Standards and Technology, 325 Broadway, Boulder, Colorado 80305, USA}
\affiliation{Department of Physics, University of Colorado, Boulder, Colorado 80309, USA}

\author{R. Srinivas}
\altaffiliation[Current address: ]{Department of Physics, University of Oxford, Oxford UK}
\affiliation{Time and Frequency Division, National Institute of Standards and Technology, 325 Broadway, Boulder, Colorado 80305, USA}
\affiliation{Department of Physics, University of Colorado, Boulder, Colorado 80309, USA}

\author{C. Arenz}
\affiliation{School of Electrical, Computer and Energy Engineering, Arizona State University, Tempe, Arizona 85287, USA}

\author{A. L. Collopy}
\affiliation{Time and Frequency Division, National Institute of Standards and Technology, 325 Broadway, Boulder, Colorado 80305, USA}

\author{L. J. Stephenson}
\affiliation{Time and Frequency Division, National Institute of Standards and Technology, 325 Broadway, Boulder, Colorado 80305, USA}
\affiliation{Department of Physics, University of Colorado, Boulder, Colorado 80309, USA}

\author{A. C. Wilson}
\affiliation{Time and Frequency Division, National Institute of Standards and Technology, 325 Broadway, Boulder, Colorado 80305, USA}

\author{\break D. J. Wineland}
\affiliation{Time and Frequency Division, National Institute of Standards and Technology, 325 Broadway, Boulder, Colorado 80305, USA}
\affiliation{Department of Physics, University of Colorado, Boulder, Colorado 80309, USA}
\affiliation{Department of Physics, University of Oregon, Eugene, Oregon 97403, USA}

\author{D. Leibfried}
\affiliation{Time and Frequency Division, National Institute of Standards and Technology, 325 Broadway, Boulder, Colorado 80305, USA}

\author{J. J. Bollinger}
\affiliation{Time and Frequency Division, National Institute of Standards and Technology, 325 Broadway, Boulder, Colorado 80305, USA}

\author{D. T. C. Allcock}
\affiliation{Time and Frequency Division, National Institute of Standards and Technology, 325 Broadway, Boulder, Colorado 80305, USA}
\affiliation{Department of Physics, University of Colorado, Boulder, Colorado 80309, USA}
\affiliation{Department of Physics, University of Oregon, Eugene, Oregon 97403, USA}

\author{D. H. Slichter}
\email{daniel.slichter@nist.gov, scburd@stanford.edu, carenz1@asu.edu}
\affiliation{Time and Frequency Division, National Institute of Standards and Technology, 325 Broadway, Boulder, Colorado 80305, USA}

\date{\today}

\begin{abstract}
We show experimentally that a broad class of interactions involving quantum harmonic oscillators can be made stronger (amplified) using a unitary squeezing protocol. While our demonstration uses the motional and spin states of a single trapped $^{25}$Mg$^{+}$ ion, the scheme applies generally to Hamiltonians involving just a single harmonic oscillator as well as Hamiltonians coupling the oscillator to another quantum degree of freedom such as a qubit, covering a large range of systems of interest in quantum information and metrology applications. Importantly, the protocol does not require knowledge of the parameters of the Hamiltonian to be amplified, nor does it require a well-defined phase relationship between the squeezing interaction and the rest of the system dynamics, making it potentially useful in instances where certain aspects of a signal or interaction may be unknown or uncontrolled.   
\end{abstract}
\maketitle

Quantum mechanical squeezing, where the uncertainty in a desired observable is reduced at the expense of increasing uncertainty in a separate, non-commuting observable, is a powerful technique for quantum-enhanced sensing and measurement~\cite{Caves1981, Yurke1986, Kitagawa1991, Wineland1992, Kitagawa1993, Pezze2018}, enabling the detection of extremely weak signals or forces~\cite{Ligo, Malnou2019, Burd2019, Davis2016, Hosten2016, Hosten2016a, Kruse2016, Gilmore2021, Backes2021, Renger2021, Jiang2022, Metelmann2022}. 
Squeezing can also be used to enhance the strength of interactions between quantum systems, for example by amplifying optomechanical or light-matter interactions relevant in quantum information science \cite{Lu2015,Lemonde2016,Zeytino2017,Qin2018,Chen2019,Leroux2018,Arenz2020,Ge2019, Ge2019b, Groszkowski2020, Li2020, Wang2022, Villiers2022}.  

The amount of enhancement in sensitivity or interaction strength that can be achieved depends not just on the strength and fidelity of the squeezing operations, but also on their timing and phase relationship relative to the other parameters in the system Hamiltonian, including the signal to be sensed or the interaction to be enhanced. In some applications, the required phases and timings for squeezing operations relative to the rest of the system dynamics may be sufficiently stable that they can be calibrated in advance. However, in other instances the parameters of the Hamiltonian may be unknown or fluctuate in time, such that na{\"i}ve application of squeezing operations can give rise to undesired ``error'' dynamics in the system in addition to the desired enhancement. 

The recently proposed scheme of ``Hamiltonian amplification'' (HA) provides a method to achieve squeezing-based enhancement of dynamics involving a quantum harmonic oscillator where parameters are fluctuating or unknown~\cite{Arenz2020, ge2020hamiltonian}. By stroboscopically applying squeezing transformations with alternating phases, errors due to unknown phase relationships are dynamically suppressed, while the desired interactions are strengthened (amplified). Crucially, the protocol does not require knowledge of the parameters of the Hamiltonian to be amplified as long as it can be written in a certain form and the timescales for the bare Hamiltonian dynamics are slow compared to the duration of applied squeezing operations. 

In this work, we experimentally realize Hamiltonian amplification using the motion of a trapped atomic ion as the quantum harmonic oscillator. As proof of principle, we use HA to demonstrate phase-insensitive amplification of coherent displacements of the ion motion, with a gain of $\approx 2$ as the relative phase of the displacement is swept across the full range of $2\pi$. We also perform phase-insensitive enhancement of laser-induced coupling between the trapped ion's motion and its internal electronic ``spin" state, where the phase of the laser interaction is not stable with respect to that of the squeezing interaction, observing an increase in the effective spin-motion coupling strength of $\approx 1.5$. 

The coupling of a quantum harmonic oscillator to another quantum system or to an external resonant driving field (representing a signal to be sensed) can be described in the interaction picture with respect to the bare harmonic oscillator Hamiltonian $\hbar\omega\hat{a}^\dagger\hat{a}$ (and all terms not involving the harmonic oscillator) by a Hamiltonian of the form
\begin{equation}
\label{eq:interactions}
\hat{H}=\hbar \Omega(\hat{\beta} \hat{a}^{\dagger}+\hat{\beta}^{\dagger}\hat{a}),
\end{equation}
where $\hat{a}$ and $\hat{a}^{\dagger}$ are the annihilation and creation operators of the quantum harmonic oscillator, $\omega$ is the harmonic oscillator frequency, and $\Omega$ characterizes the interaction strength with either an external driving field\textemdash in which case $\hat{\beta}$ is just a complex number\textemdash or a second quantum system, where $\hat{\beta}$ is an operator for that system.
For example, in the case where $\hat{\beta}=\hat{\sigma}^-$, the lowering operator for a spin-1/2 system, equation~(\ref{eq:interactions}) is a Jaynes-Cummings-type coupling. However, in general the operator $\hat{\beta}$ describing the second quantum system does not need to be specified for the HA procedure to work, which makes the method amenable to amplifying interactions in a wide range of systems.

An external resonant driving field with constant amplitude applied to the harmonic oscillator for a duration $t_d$ will cause a coherent displacement from the time evolution $\exp(-i\hat{H}t_d/\hbar)=\hat{D}(\alpha)$, where $\hat{D}(\alpha)\equiv\exp(\alpha\hat{a}^{\dagger}-\alpha^{*}\hat{a})$ is the displacement operator and $\alpha=-i\Omega\beta t_d$, where we have dropped the hat from $\beta$ since here it is just a complex number and not an operator. Measuring the displacement $\alpha$ constitutes sensing of the external field during $t_d$. If $|\alpha|\ll 1$ (sensing a very weak signal), amplifying the displacement will enable it to be more easily resolved with respect to the zero-point fluctuations of the harmonic oscillator, which set the noise floor for measuring displacements. The desired amplification can be achieved by applying appropriate squeezing and anti-squeezing operations before and after the sensing period according to the identity~\cite{Nieto1997, Burd2019}
\begin{equation}
\label{eq:phase_sens}
\hat{S}^{\dagger}(\xi)\hat{D}(\alpha)\hat{S}(\xi)=\hat{D}(\alpha_{amp}).
\end{equation}
\noindent Here $\hat{S}(\xi)=\exp((\xi^{*} \hat{a}^{2}-\xi \hat{a}^{\dagger 2})/2)$ is the squeezing operator with the complex squeezing parameter $\xi=r\exp(i\theta)$, and the amplified displacement is given by $\alpha_{amp}=\alpha \cosh(r)+\alpha^{*}e^{i\theta}\sinh(r)$; $r$ is taken to be real and positive. The angle $\theta$ describes the quadrature of the harmonic oscillator that is squeezed by the application of $\hat{S}(\xi)$. In harmonic oscillator phase space, viewed in the frame rotating at $\omega$, the squeezed axis makes an angle of $\theta/2$ with respect to the real axis.

\begin{figure}[t]
\includegraphics[width=0.48\textwidth]{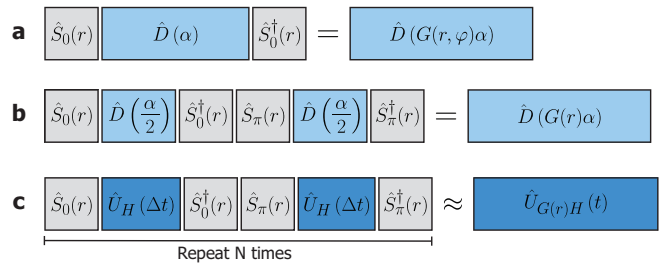}
\centering
\caption{\textbf{Pulse sequences for different quantum amplification schemes using unitary squeezing.} \textbf{a}, Pulse sequence for the phase-sensitive amplification scheme used in Ref.~\cite{Burd2019}. \textbf{b}, Pulse sequence for Hamiltonian amplification of a coherent displacement. \textbf{c}, Pulse sequence for Hamiltonian amplification of a general interaction as in equation~\eqref{eq:unitaryamp}, which must be appropriately Trotterized. 
}
\label{fig:pulse_seq}
\end{figure}

The pulse sequence corresponding to equation~(\ref{eq:phase_sens}) is shown in Fig.~\ref{fig:pulse_seq}a. Both the magnitude and the phase of the amplitude gain $G\equiv\alpha_{amp}/\alpha$ depend on the phase relationship between the displacement and the squeezing, with $G(r,\varphi)=\cosh(r)+e^{i(\theta-2\varphi)}\sinh(r)$, where $\varphi=\mathrm{arg}(\alpha)$ is the phase of the displacement in the previously described rotating frame of the harmonic oscillator. Achieving maximum gain requires that $\theta-2\varphi= 2 n \pi$ for integer $n$, giving $G=e^r$; fluctuations in this phase difference will cause corresponding gain fluctuations, and for some values of phase the displacement will even be de-amplified, with $|G|<1$. Depending on the nature of the physical system, it may be challenging to stabilize $\theta$ and $\varphi$ with respect to each other. 

\begin{figure}[t]
\includegraphics[width=0.48\textwidth]{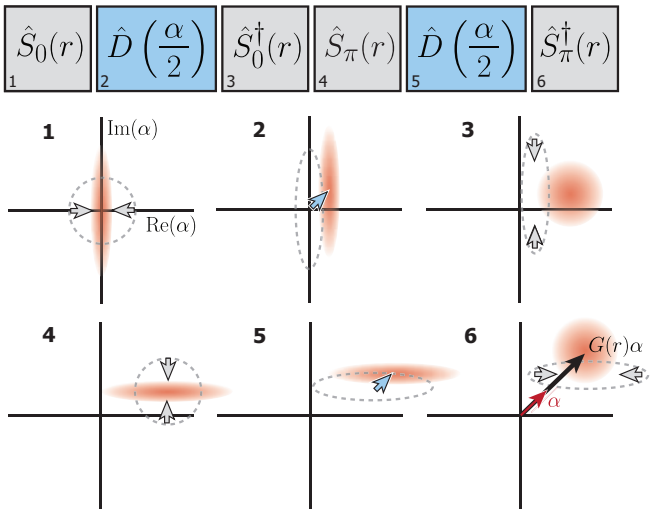}
\centering
\caption{ \textbf{Schematic of Hamiltonian amplification of a coherent displacement.} We plot a series of cartoon schematic representations of the motional phase space distribution, before (dotted gray outline) and after (orange) each numbered pulse. The arrows indicate squeezing (gray) and displacement (blue) operations, respectively. In the final panel, the red arrow shows the total unamplified displacement, while the black arrow shows the amplified displacement, which points in the same direction but is larger by a factor $G(r)$.
}
\label{fig:pulse-blobs}
\end{figure}

We seek to perform phase-independent amplification, where $G$ is independent of $\theta$ and $\varphi$. For amplification of displacements, this can be achieved by dividing the displacement into two steps, $\hat{D}(\alpha)=\hat{D}(\alpha/2)\hat{D}(\alpha/2)$, and amplifying each step individually with squeezing and anti-squeezing operations. Crucially, the second squeeze/anti-squeeze pair amplifies the harmonic oscillator quadrature that is 90 degrees out of phase with respect to the quadrature amplified by the first pair; mathematically, the value of $\theta$ is increased by $\pi$ for the second squeeze/anti-squeeze pair. Without loss of generality, we can take the values of $\theta$ to be $0$ and $\pi$ for the first and second squeeze/anti-squeeze pairs, respectively. The sequence of operations, shown in Fig.~\ref{fig:pulse_seq}b and also in Fig.~\ref{fig:pulse-blobs}, is now
\begin{align}
\label{eq:phasindependentamp}
\hat{D}(\alpha_{amp})=\hat{S}_{\pi}^{\dagger}(r)\hat{D}(\alpha/2)\hat{S}_{\pi}(r)\hat{S}_{0}^{\dagger}(r)\hat{D}(\alpha/2)\hat{S}_{0}(r),	
\end{align}
\noindent where the subscripts on $\hat{S}$ indicate the value of $\theta$. We note the identity $\hat{S}_0^{\dagger}(r)=\hat{S}_\pi(r)$, such that the second pair of squeezing pulses are identical to the first pair but with reversed order. Using the identity in equation~(\ref{eq:phase_sens}), and the fact that the product of two displacement operators is just another displacement operator times a complex phase, we find that $\hat{D}(\alpha_{amp})=\exp(\cosh(r)(\alpha \hat{a}^{\dagger}-\alpha^{*}\hat{a}))$, up to a global phase. As such, the sequence \eqref{eq:phasindependentamp} yields a gain of
\begin{align}
\label{eq:phaseindepshort}
G(r)=\cosh(r),
\end{align}	
independent of the phases $\theta$ and $\varphi$. The price for phase insensitivity is a reduction in the gain $G$ relative to the maximum achievable with the phase-sensitive scheme in equation~(\ref{eq:phase_sens}); for $r\gtrsim 1$ the gain is reduced by a factor of $\approx 2$.

The situation is somewhat more complicated when squeezing is used to amplify general interactions of the form of equation~(\ref{eq:interactions}), where $\hat{\beta}$ is an operator and not a complex number. Consider a time evolution modified through instantaneous squeezing and anti-squeezing, performed sequentially along two orthogonal quadratures in phase space according to  
\begin{align}
\label{eq:unitaryamp}
\hat{U}(t)=\hat{S}_{\pi}^{\dagger}(r)\hat{U}_{H}(t/2)\hat{S}_{\pi}(r)\hat{S}_{0}^{\dagger}(r)\hat{U}_{H}(t/2)\hat{S}_{0}(r),	
\end{align}
where $\hat{U}_{H}(t)=\exp(-i \hat{H} t/\hbar)$ describes the time evolution for a time interval $t$ under the Hamiltonian in equation~(\ref{eq:interactions}). In general equation~(\ref{eq:unitaryamp}) yields an infinite power series in $t$, giving ``error'' terms in addition to the desired amplified interaction. To address this, we can Trotterize the interaction~\cite{Trotter1959,Suzuki1976}, splitting the total interaction time $t$ into small segments $\Delta t=\frac{t}{2N}$ with $N\in\mathbb N$ such that $\hat{U}(t)=\left[\hat{U}(2\Delta t)\right]^{N}$ and repeating the sequence in equation~(\ref{eq:unitaryamp}) $N$ times. For sufficiently large $N$ (or equivalently, sufficiently small $\Delta t$), we can neglect higher-order terms in $\Delta t$ and find 
\begin{align}
\label{eq:amplifieddyn}
\hat{U}_{G(r)H}(t)\approx \exp(-i\cosh(r) \hat{H} t/\hbar).	
\end{align}
In this regime, the system dynamics are approximately governed by an amplified Hamiltonian $\hat{H}_\mathrm{amp}=G(r)\hat{H}$ with phase-independent gain $G(r)=\cosh(r)$, thus the name ``Hamiltonian amplification''~\cite{Arenz2020}. We remark here that while the sequence given in equation~(\ref{eq:unitaryamp}) rests on the assumption that the evolution generated by the interaction $\hat{H}$ can be neglected during squeezing, this assumption is in general not necessary for HA to work. Large, smooth, high-frequency modulations of the squeezing parameter can be used instead to achieve amplification of interactions in systems where this assumption cannot be met \cite{Arenz2020}. According to the Trotter formula, the dynamics in equation~(\ref{eq:amplifieddyn}) become exact in the limit of $N\to \infty$~\cite{Arenz2020, ge2020hamiltonian}. For finite $N$, it is generally challenging to bound the Trotter error for systems with infinite dimensional Hilbert spaces \cite{TrotterConvergence}. However, a rough estimate for the required fineness of the Trotterization is given by the scaling of the second order term in $\Delta t$, yielding $\Delta t\ll\left[\Omega\sqrt{\sinh(2r)}\right]^{-1}$, ignoring the dependence of the estimate on the operators in this second order term~\cite{Arenz2020}.

The experiment uses a single $^{25}$Mg$^{+}$ ion confined in a surface-electrode ion trap~\cite{Seidelin2006} containing current-carrying electrodes for producing strong near-field magnetic field gradients at the ion position, 30 $\mu$m above the electrode plane. Details of the experimental apparatus have been presented elsewhere~\cite{Srinivas2019, Burd2019, Burd2021, Srinivas2021}. The harmonic oscillator degree of freedom is a radial mode of ion motion with frequency $\omega/2\pi\sim7$\, MHz, whose state we describe in the number state basis $\ket{n}$. The ion motion is cooled near its ground state (mean phonon occupation $\bar{n}=0.06$) by Doppler cooling with laser light at 280 nm followed by resolved sideband cooling~\cite{Monroe1995}. Preparation and analysis of the motional quantum state rely on coupling the motion to a qubit encoded in the internal (electronic) states of the ion via motional sideband transitions~\cite{Wineland1998, Leibfried2003}. We choose the $\ket{\downarrow}\equiv\ket{F=3,m_F=1}$ and $\ket{\uparrow}\equiv\ket{F=2,m_F=1}$ states of the $^2S_{1/2}$ electronic ground state hyperfine manifold as qubit states (qubit frequency $\omega_q/2\pi=1.686$ GHz with a quantization magnetic field of 21.3 mT), and implement sideband transitions for cooling and motional state analysis using near-field magnetic field gradients oscillating at $\omega_q\pm \omega$~\cite{Ospelkaus2008, Ospelkaus2011}. Coherent displacements of the motional state are performed by applying a resonant electric potential at $\omega$ to a trap electrode\textemdash producing a corresponding oscillating, spatially uniform electric field at the ion\textemdash for a fixed duration~\cite{Wineland1998,Burd2019}. Squeezing and anti-squeezing of the motional state is achieved by applying an electric potential to the trap rf electrodes at frequency $2\omega$, which parametrically modulates the confining potential~\cite{Heinzen1990,Burd2019, Burd2021}. The magnitude $r$ of the squeezing parameter can be adjusted either by varying the squeezing pulse duration $t_{s}$ or by changing the amplitude of the parametric drive and thus the parametric modulation strength $g$ (see Methods). The electronic waveforms used to implement the resonant and parametric drives are generated using phase-synchronized direct digital synthesizers (DDSs). By adjusting the relative phase of these waveforms, we can arbitrarily control the squeezing angle $\theta$ and the displacement direction $\varphi$ in motional phase space. For all sequences, we characterize the initial and final motional states using a magnetic gradient-based blue sideband (BSB) analysis pulse that couples the motional state to the internal qubit states of the ion. By measuring the qubit state populations for varying BSB pulse durations, we can extract the motional Fock state populations (the diagonal elements of the motional state density matrix in the Fock basis; see Methods)~\cite{Meekhof1996}. We then characterize the measured states by fitting the extracted Fock state populations to parameterized models of Fock state populations corresponding to either coherent states or displaced squeezed states (see Extended Data Fig.~\ref{fig:calibrations} and Methods). We verify the phase coherence of squeezed states by anti-squeezing them and verifying that the motion returns to near the ground state (see Methods)~\cite{Burd2019}.

\begin{figure}[tb]
\includegraphics[width=0.5\textwidth]{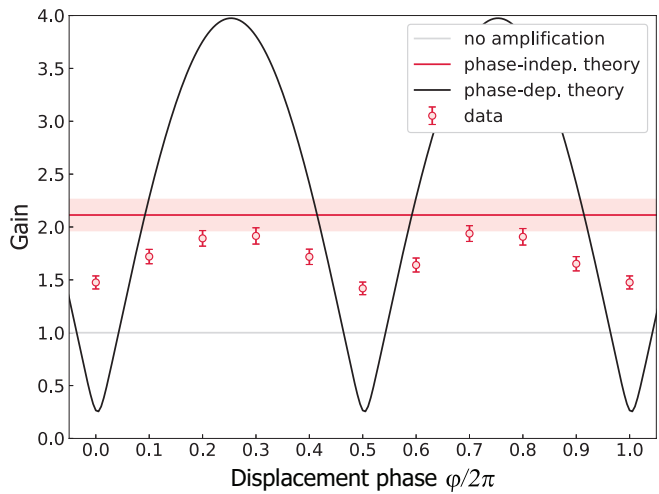}
\centering
\caption{ \textbf{HA gain versus displacement phase.} Experimental data are shown as red circles, with error bars denoting 68\% confidence intervals. The red line (light red band) shows the theoretical prediction (68\% confidence interval) for the HA gain $G=\cosh(r)$ given the experimentally determined value of of $r=1.38(8)$. The theoretical phase-dependent gain of the scheme in Fig.~\ref{fig:pulse_seq}a given the same $r$ is shown for comparison as the black line. The gray line denotes $G=1$. The data point plotted at $\varphi/2\pi=1$ is a copy of the data point at $\varphi/2\pi=0$.}
\label{fig:amp-disp-data}
\end{figure}

We first perform HA of coherent displacements, a method that could be used for quantum-enhanced sensing of unknown weak resonant fields, using the pulse sequence in Fig. \ref{fig:pulse_seq}b. The squeezing and displacement operations have durations of 2.75~$\mu$s and 7~$\mu$s, respectively. To characterize the phase sensitivity of the protocol, we perform the same amplification sequence with fixed $\theta$ for ten different values of $\varphi$ uniformly spaced to cover the interval $[0, 2\pi)$, and measure the resulting gain. We calibrate the displacement strength by setting $r=0$ (replacing squeezing with a delay of equivalent duration), measuring a coherent state amplitude $\alpha_i=0.55(2)$. The squeezing strength $r=1.38(8)$ is extracted by squeezing the initial state (approximately the ground state) and fitting the resulting Fock state populations (see Methods). With the squeezing turned on in the HA sequence, we measure a mean final displacement amplitude (averaged over all displacement phases) of $\alpha_{amp}=0.979(3)$; data are shown in Fig.~\ref{fig:amp-disp-data}. The corresponding mean gain is $G=1.77(5)$, in reasonable agreement with the theoretically expected gain of $\cosh(r)=2.11(15)$. We see some residual phase dependence of the gain due to imperfect calibration of the strengths and phases of the squeezing pulses, as well as drifts in the squeeze drive (see Methods). We can also perform this HA by subdividing $\alpha_i$ more finely and performing a generalized HA sequence with $N$ rounds as shown in Fig.~\ref{fig:pulse_seq}c. No Trotterization error occurs when amplifying a coherent displacement, so we can choose any value of $N$. We find that increasing $N$ improves the phase insensitivity at the cost of decreased overall gain (see Extended Data Fig.~\ref{fig:amp-disp-n3}), which we believe is due to amplified motional decoherence from the additional squeezing pulses, as well as imperfect pulse calibrations.

Next, we demonstrate the use of HA to enhance the coupling between a quantum harmonic oscillator and a qubit, characterized by state transition operators $\hat{\sigma}^\pm\equiv\frac{1}{2}(\hat{\sigma}_x\mp i \hat{\sigma}_y)$ constructed from Pauli operators $\hat{\sigma}_x$ and $\hat{\sigma}_y$. In our experiment, we can drive a red sideband (RSB) transition on the ion to realize the Jaynes-Cummings (JC) Hamiltonian

\begin{equation}
\label{eq:JC}
\hat{H}_{JC}=\hbar \Omega(\hat{\sigma}^- \hat{a}^{\dagger}+\hat{\sigma}^+\hat{a})\, 
\end{equation}

\noindent where the qubit states $\ket{\tilde{\downarrow}}\equiv{^2S}_{1/2}\ket{F=3,m_F=3}$ and $\ket{\tilde{\uparrow}}\equiv{^2S}_{1/2}\ket{F=2,m_F=2}$ have energy splitting $\hbar\widetilde{\omega}_q$, with $\widetilde{\omega}_q/2\pi = 1.326$~GHz, and the harmonic oscillator is a motional mode of the ion. The qubit-harmonic oscillator coupling is realized with a laser-driven stimulated Raman transition~\cite{Wineland1998}, using two non-copropagating laser beams at 280 nm with a frequency difference of $\widetilde{\omega}_q-\omega$. We choose a different pair of hyperfine states ($\ket{\tilde{\uparrow}}$ and $\ket{\tilde{\downarrow}}$) as the qubit due to experimental constraints on the frequency difference of the Raman laser beams. The phase of the RSB interaction is determined by the (unstabilized) phase of the optical beat note of the two laser beams at the ion, which is sensitive to differential optical path length fluctuations for the two beams and can vary over a substantial fraction of $2\pi$ between successive experimental trials. Amplifying these dynamics with HA again demonstrates that HA is phase-independent, since otherwise individual experimental shots would exhibit widely varying gain, including gains below 1 (de-amplification)~\cite{Burd2021}. 

\begin{figure}[t]
\includegraphics[width=0.52\textwidth]{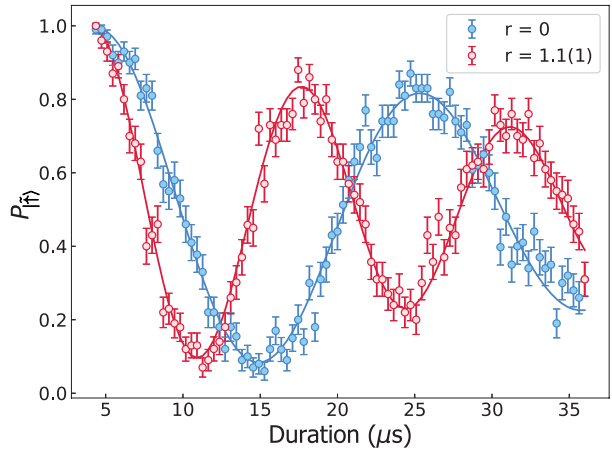}
\centering
\caption{\textbf{Hamiltonian amplification of the Jaynes-Cummings (JC) interaction.} Population in $\ket{\tilde{\uparrow}}$ as a function of the JC interaction duration for a Hamiltonian amplification sequence with $N=6$ (the duration of the squeezing pulses is not included). The JC interaction induces Rabi oscillations between the $\ket{\tilde{\uparrow}}\ket{n=0}$ and $\ket{\tilde{\downarrow}}\ket{n=1}$ states with a rate $\Omega$. Increasing the squeezing parameter from $r=0$ to $r=1.1$ increases $\Omega$ by a factor of 1.56(2). The solid lines are fits to exponentially decaying sinusoids. Error bars indicate 68 \% confidence intervals.} 
\label{fig:RHA}
\end{figure}

  Amplification of $\hat{H}_{JC}$ is achieved by interleaving a sequence of laser-induced red sideband pulses with squeezing pulses as shown in Fig.~\ref{fig:pulse_seq}c. Since $[\hat{H}_{JC},\hat{S}(\xi)]$ is an operator and not a complex number, a Trotterized HA sequence with sufficiently high $N$ is required for faithful amplification of the Hamiltonian according to equation~(\ref{eq:amplifieddyn}); experimentally, we find that $N=6$ gives qualitatively small ``error'' terms from Trotterization. These experiments used longer, weaker squeezing pulses which help reduce the residual effects of imperfections in the squeezing, but also extend the interaction durations, increasing motional decoherence.
  The JC interaction induces Rabi oscillations between the $\ket{\tilde{\uparrow}}\ket{n=0}$ and $\ket{\tilde{\downarrow}}\ket{n=1}$ states with Rabi frequency $\Omega$. Figure~\ref{fig:RHA} shows the population in the $\ket{\tilde{\uparrow}}$ state as a function of the total duration of the Trotterized laser sideband pulses applied (the duration of the squeezing pulses, which is 144 $\mu$s total for $N=6$, is not included). We measure an increase in $\Omega$ by a factor of 1.56(2) (determined by fitting to an exponentially decaying sinusoid) under HA with a squeezing parameter $r=1.1(1)$, relative to the case without squeezing $(r=0)$, as shown in Fig.~\ref{fig:RHA}. For comparison, the theoretically expected gain from HA is $\cosh r=1.7(1)$.

We calculate theoretically that dephasing of the motional degrees of freedom is sped up by HA along with the RSB interaction (see Methods). However, non-unitary processes of the two-level system, such as off-resonant Raman scattering or qubit dephasing, are unaffected by HA, making this technique potentially valuable in practice for systems where decoherence of the two-level system dominates over motional decoherence. Developing a valid description of the open quantum harmonic oscillator system driven by infinitely strong and fast parametric controls is challenging and will be the subject of future studies. 

In summary, we have implemented the proposed method of Hamiltonian amplification~\cite{Arenz2020} in a trapped-ion system; the method can be used in any physical system where fast unitary squeezing can be implemented. For example, low-noise, phase-insensitive amplification of coherent displacements of the electromagnetic field in a cavity could improve the sensitivity of axion dark matter detectors~\cite{Malnou2019,Caves2019,Backes2021}. In quantum information and simulation platforms where boson-mediated interactions are essential for generating entanglement, Hamiltonian amplification could potentially mitigate the impact of qubit decoherence~\cite{Burd2021}.

\begin{acknowledgments}
	We thank A. L. Carter and A. D. Brandt for helpful comments on the manuscript. C.A. conceived of the HA scheme and developed the theory; S.C.B. and H.M.K. performed the experiments and analyzed the data, assisted by R.S., A.L.C., L.J.S., and D.H.S.; the manuscript was written by D.H.S., C.A., S.C.B., and H.M.K., with input from all authors; D.H.S. and D.T.C.A. supervised the work with support from D.L., A.C.W., J.J.B., and D.J.W.; A.C.W., D.L., D.J.W., and D.H.S. secured funding for the work; all authors contributed to scientific discussions and planning. This work was supported by NIST (https://ror.org/05xpvk416).
\end{acknowledgments}

\bibliographystyle{naturemag}
\bibliography{HA}

\setcounter{figure}{0}
\renewcommand{\figurename}{Extended Data Fig.}

\section{Methods}

\subsection{Hamiltonian Amplification in the presence of bosonic dephasing}\label{sec:HAdephasing}
We consider a composite system described by the Hamiltonian $\hat{H}=\hbar \Omega(\hat{\beta} \hat{a}^{\dagger}+\hat{\beta}^{\dagger}\hat{a})$. We assume that the quantum harmonic oscillator represented by annihiliation and creation operators $\hat{a}$ and $\hat{a}^\dagger$ is subject to dephasing described by the dissipator \cite{Turchette2000} 
\begin{align}
\mathcal{D}_{a}(\cdot)=-\Gamma \left[ \hat{a}^{\dagger}\hat{a},[\hat{a}^{\dagger}\hat{a},(\cdot)] \right],
\end{align}
where non-unitary processes of the system ($\hat{\beta}$) interacting with the quantum harmonic oscillator are described by a dissipator $\mathcal D_{\beta}$, which we do not specify further. The total system (i.e., coherent and dissipative part) is described by a Lindbladian of the form 
\begin{align}
\label{eq:LindbladmaterEq}
\mathcal L(\cdot)=-\frac{i}{\hbar}[\hat{H},(\cdot)]+\mathcal D_{a}(\cdot)+\mathcal D_{\beta}(\cdot). 
\end{align}
The system dynamics at time $t$ is then given by the completely positive and trace preserving map $\Lambda_{t}=\exp(\mathcal L t)$. Now, we modify the dynamics by alternating in time intervals $\Delta t=t/2N$ between instantaneously squeezing in two quadratures of the harmonic oscillator, described by the unitary maps $\mathcal S_{0}(\cdot)=\hat{S}_{0}^{\dagger}(\cdot)\hat{S}_{0}$ and $\mathcal S_{\pi}(\cdot)=\hat{S}_{\pi}^{\dagger}(\cdot)\hat{S}_{\pi}$, respectively, to realize the Hamiltonian amplification protocol. In the limit of infinitely fast alternation between squeezing quadratures, the dynamics are given by 
\begin{align}
\lim_{N\to\infty}\left(\mathcal S_{\pi}^{\dagger}\Lambda_{\Delta t}\mathcal S_{\pi}\mathcal S_{0}^{\dagger}\Lambda_{\Delta t}\mathcal S_{0}\right)^{N}=e^{\bar{\mathcal L}t},
\end{align}
where the effective Lindbladian $\bar{\mathcal L}$ reads \cite{PhysRevA.92.022102},
\begin{align}
&\bar{\mathcal L}(\cdot)=\frac{1}{2}\sum_{j\in\{0,\pi\}} \left( -i \left[ \mathcal S_{j}(H),(\cdot) \right] +\mathcal S_{j}^{\dagger}\mathcal D_{a} \left( \mathcal S_{j}(\cdot) \right) \right)+\mathcal D_{\beta}(\cdot). 
\end{align}
After some algebra, and using standard properties of the squeezing operator, we find 
\begin{align}
\label{eq:effectiveLind}
\bar{\mathcal L}(\cdot)&=-i\cosh(r)[\hat{H},(\cdot)]+\cosh^{2}(2r)\mathcal D_{a}(\cdot)\nonumber \\
&+\frac{1}{4}\sinh^{2}(2r) \left[ \hat{a}^{\dagger 2}+\hat{a}^{2},[\hat{a}^{\dagger 2}+\hat{a}^{2},(\cdot)] \right] \nonumber \\
&+\mathcal D_{\beta}(\cdot).
\end{align}
Thus, while the coherent part is amplified by a factor $\lambda=\cosh(r)$, the dephasing  $\mathcal D_{a}$ of the quantum harmonic oscillator is amplified by a larger factor $\cosh^{2}(2r)$. Moreover, we see that $\mathcal D_{a}$ is not just amplified, but also modified under HA, with an additional dephasing-type term appearing in the second line of equation \eqref{eq:effectiveLind} for the effective Lindbladian. We note that the non-unitary process generated by $\mathcal D_{\beta}$, which acts on the system described by $\hat{\beta}$ and $\hat{\beta}^\dagger$ that interacts with the quantum harmonic oscillator, is not amplified or modified by HA. For example, if $\hat{\beta}=\hat{\sigma}^-$ such that $\hat{H}$ is the Jaynes-Cummings interaction, then $\mathcal D_{\beta}(\cdot)$ describes decoherence of the qubit coupled to the harmonic oscillator. If this decoherence is dominated by spontaneous off-resonant photon scattering due to the Raman beams generating the Jaynes-Cummings interaction, then the HA process may increase the desired interaction strength without increasing the decoherence rate, thus improving the fidelity of the interactions. However, future work is needed to assess the validity of describing the open system dynamics in the presence of HA (i.e., an open system subject to infinitely strong and fast applied periodic parametric controls) through a Lindblad type master equation of the form \eqref{eq:LindbladmaterEq}. In fact, in a similar setting it has been argued that such as description of the open system dynamics can fail~\cite{PhysRevA.92.022102}.

\subsection{Motional state analysis}

\begin{figure*}[tb]
\includegraphics[width=0.93\textwidth]{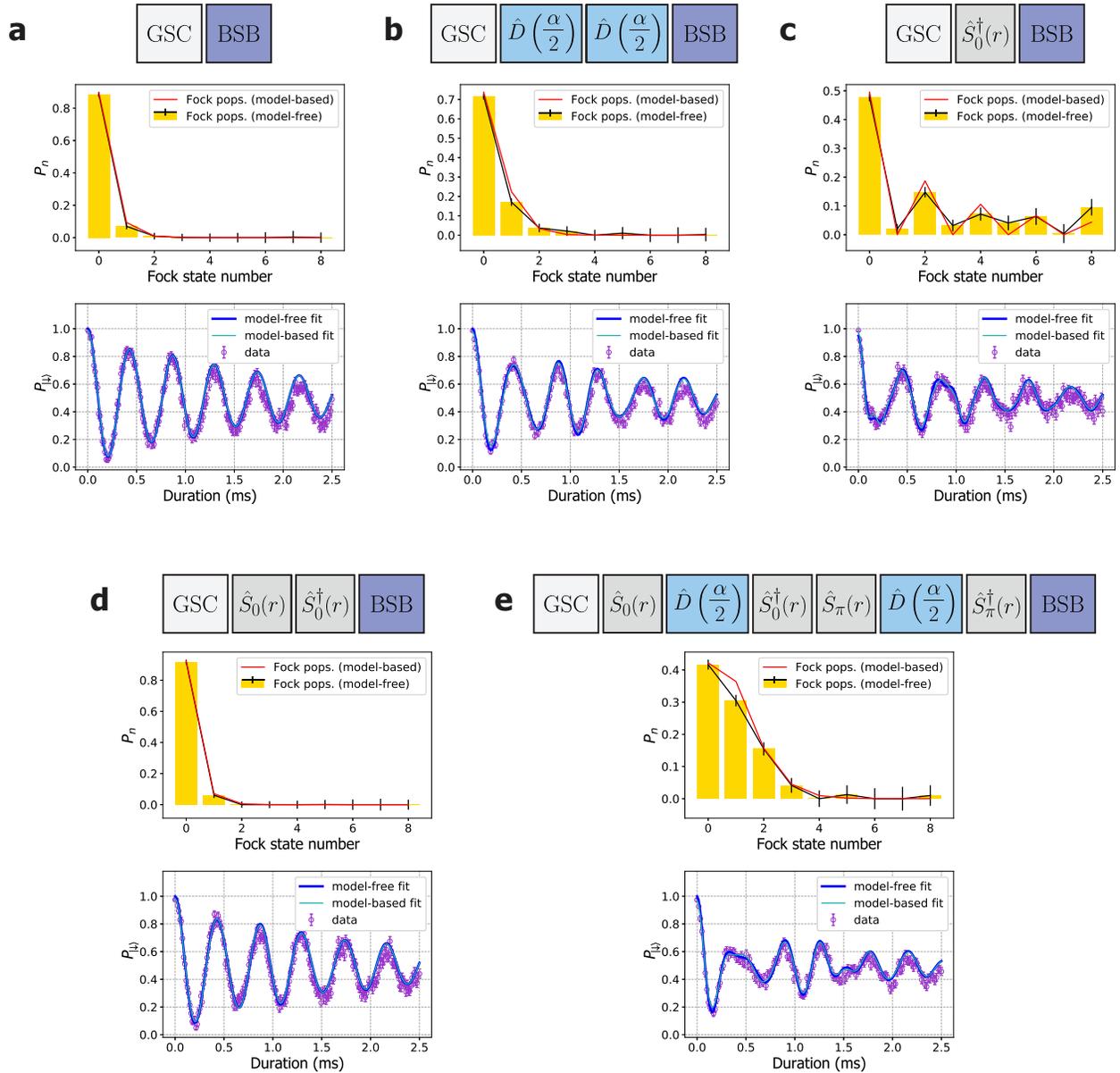}
\centering
\caption{\textbf{Pulse sequences and calibration data.} GSC: ground state cooling, BSB: blue sideband analysis pulse. Displacement and squeezing operators are the same as in Fig.~\ref{fig:pulse_seq}. The lower plot in each panel shows the measured population in $\ket{\downarrow}$ (purple circles) as a function of blue sideband (BSB) analysis pulse duration for various initial motional states. The dark blue lines show the result of model-free fits to the data from equation~(\ref{eq:BSB_flopping}) as described in the text, and the light blue lines show the model-based fits used to extract the various parameters ($r$, $\bar{n}$, $\alpha$). Both the Fock state plots and BSB flops demonstrate good agreement between the model-based and model-free fits, validating our choice of fit models. The upper plot in each panel shows the corresponding Fock state populations extracted by fitting to the BSB oscillations. The red lines show the best-fit Fock state populations from a model-based fit to the BSB oscillations, while the yellow bars show the Fock state populations extracted from a model-free fit. The vertical black lines show 68\% confidence intervals on the model-free fitted populations. Diagonal black lines connect the tops of the yellow bars to guide the eye and highlight deviations between the model-based and model-free fitted populations.  \textbf{a}, Ground state cooling calibration. We find a mean occupation of $\bar{n}=0.06(1)$. \textbf{b}, Displacement calibration. Following ground state cooling, the motional state is displaced by an applied resonant excitation pulse split into two 3.5 $\mu$s segments (7 $\mu$s total), yielding $\alpha_i=0.55(2)$. \textbf{c}, Squeezing calibration. Following ground state cooling, the parametric drive is applied for 2.75 $\mu$s to generate a squeezed state of motion. \textbf{d}, Anti-squeezing calibration and squeezing coherence verification. The ion is cooled to its motional ground state, the parametric drive is applied for 2.75 $\mu$s, and then the parametric drive is applied for an additional 2.75 $\mu$s with a $180^\circ$ phase shift. Fitting the resulting state to a model for a thermal state, we find $\bar{n}=0.09(1)$. \textbf{e}, Hamiltonian amplification experiment. Final motional state after HA of a coherent displacement with $N=1$, using two displacment pulses of 3.5 $\mu$s duration and four squeezing/anti-squeezing pulses of 2.75 $\mu$s duration. The data in Fig.~\ref{fig:amp-disp-data} are produced by comparing the calibrated unamplified displacement as seen in \textbf{b} to the displacement extracted in \textbf{e} for different relative phases between the displacement and squeezing pulses.}
\label{fig:calibrations}
\end{figure*}

Analysis of the motional state populations is accomplished by preparing the desired motional state with the qubit in the $\ket{\downarrow}$ state, then applying a magnetic gradient-based BSB interaction to couple the motional state to the qubit states. This maps information about the motional state onto the qubit state; by measuring the population in the qubit $\ket{\downarrow}$ state using fluorescence detection for varying durations of the BSB interaction, we can extract the state populations in the motional Fock basis. For an arbitrary motional state the measured population in $\ket{\downarrow}$ is given by~\cite{Meekhof1996, Kienzler2015, Burd2019}

\begin{equation}
P_{\downarrow}(t)=\frac{1}{2} \left[ 1+\sum_{n=0}^{\infty}P_{n}e^{-\gamma\sqrt{n+1}t}\cos \left( \Omega_{SB}\sqrt{n+1}t \right) \right]
\label{eq:BSB_flopping}
\end{equation}

\noindent
where $n$ denotes the oscillator Fock state, $t$ is the duration of the sideband interaction, $\Omega_{SB}$ is the Rabi frequency of the sideband (for the $\ket{\downarrow}\ket{n=0}\leftrightarrow \ket{\uparrow}\ket{n=1}$ transition), and $\gamma$ is a phenomenological decay constant. We note that Ref.~\onlinecite{Meekhof1996} uses a multiplier of $(n+1)^{0.7}$ for $\gamma$ while Refs.~\onlinecite{Kienzler2015, Burd2019} use a multiplier of $(n+1)^{0.5}$ as is done here; this choice has a small effect on goodness of fit but does not affect the extracted Fock state populations. The factor of $\sqrt{n+1}$ multiplying $\Omega_{SB}$ assumes the system is in the Lamb-Dicke regime, meaning here that the microwave magnetic field gradient used to perform the sideband operation is homogeneous over the spatial extent of the ion motional wavepacket~\cite{Wineland1998}. This is an excellent approximation for our system~\cite{Burd2019}. The Fock state probability distribution of the initial motional state is given by $\{P_{n}\}$. We fit the measured traces of $P_{\downarrow}(t)$ to extract the values of $\{P_n\}$. For specific classes of motional states the $\{P_n\}$ are related in a parameterized way; for example, a coherent state parameterized by the amplitude $\alpha$ can be described by the model~\cite{Gerry2005}

\begin{equation}
P_{n}=\frac{e^{-|\alpha|^{2}}|\alpha|^{2n}}{n!},
\label{eq:coherent}
\end{equation}

\noindent while the squeezed vacuum state parameterized by complex squeezing parameter $\xi=re^{i\theta}$ can be described by the model~\cite{Gerry2005}

\begin{equation}
P_{n}=\frac{(\tanh r)^{n}}{\cosh r}\frac{n!}{2^n}\left (\frac{1}{(n/2)!}\right )^{2},
\label{eq:squeezed_state_Pn}
\end{equation}

\noindent for even $n$, with $P_n=0$ for odd $n$. Extended Data Fig.~\ref{fig:calibrations} shows example traces of the $\ket{\downarrow}$ population versus BSB duration, along with extracted motional state populations in the Fock basis. The upper plots in each panel show the extracted populations for both ``model-free'' fits, fitting to equation~(\ref{eq:BSB_flopping}) to determine the $P_n$ while only constraining that $\sum_{n=0}^\infty{P_n} = 1$, as well as ``model-based'' fits, where the $P_n$ are all related according to one of the parameterized models described above. The lower plots in each panel show the experimental data along with the corresponding model-based and model-free fit curves.

\begin{figure}[tb]
\includegraphics[width=0.5\textwidth]{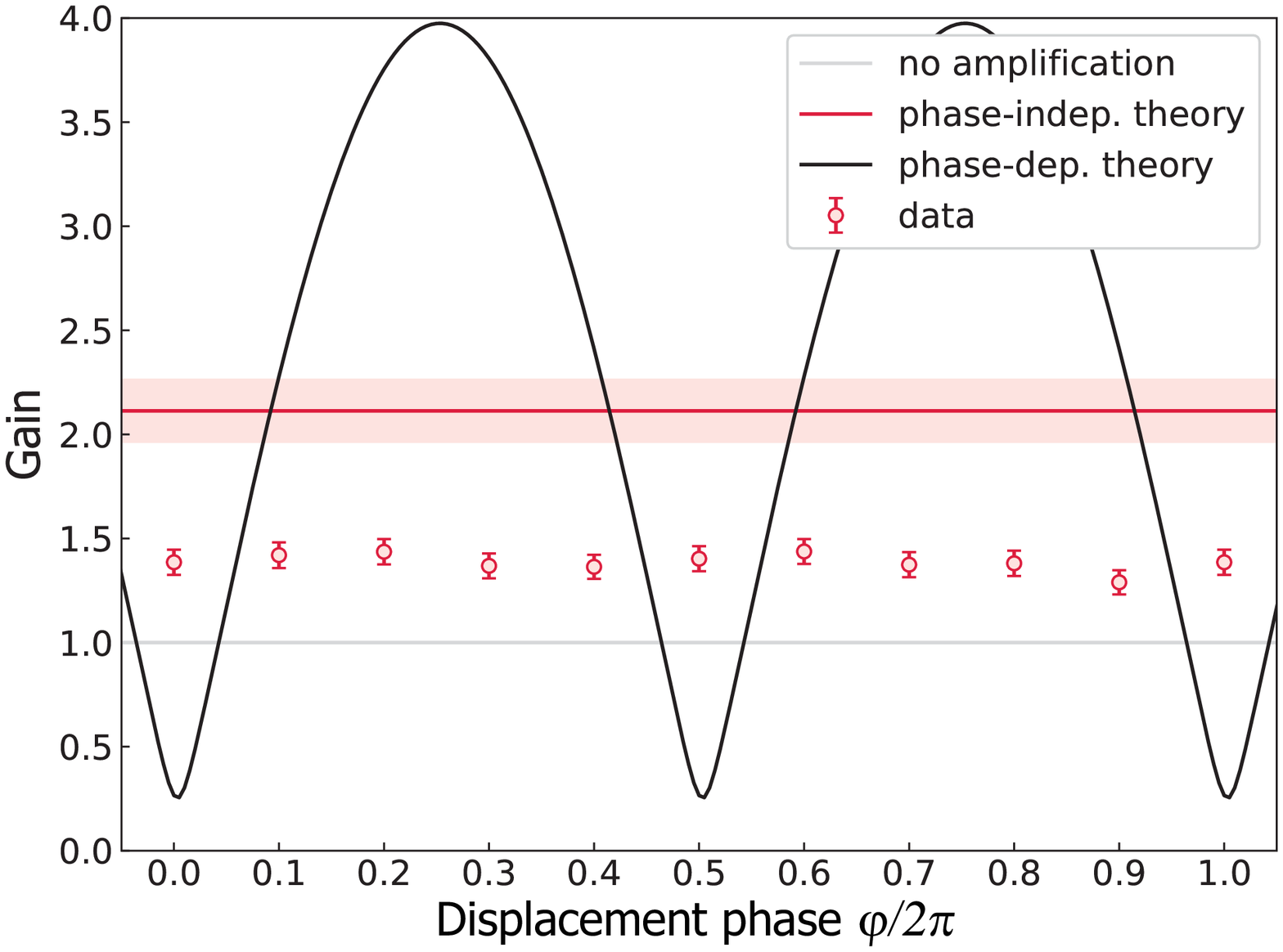}
\centering
\caption{ \textbf{HA gain versus displacement phase for $N=3$.} This plot shows the same experiment as in Fig.~\ref{fig:amp-disp-data} but with $N=3$ rather than $N=1$. This means the displacement is split into six equal pieces rather than two, which due to experimental imperfections results in decreased amplification but improved phase insensitivity. Experimental data are shown as red circles, with error bars denoting 68\% confidence intervals. The red line (light red band) shows the theoretical prediction (68\% confidence interval) for the HA gain $G=\cosh(r)$ given the experimentally determined value of of $r=1.38(8)$. The theoretical phase-dependent gain of the scheme in Fig.~\ref{fig:pulse_seq}a given the same $r$ is shown for comparison as the black line. The gray line denotes $G=1$. The data point plotted at $\varphi/2\pi=1$ is a copy of the data point at $\varphi/2\pi=0$.}
\label{fig:amp-disp-n3}
\end{figure}

The $r$ value reported in the main text is a mean of two squeezing calibrations, one measured before the data were taken and a second one measured after, as the actual experimental squeezing strength drifts more than is captured in the fit uncertainty. The value calibrated just before the data in Fig.~\ref{fig:amp-disp-data} and Extended Data Fig.~\ref{fig:amp-disp-n3} were taken is $r=1.42(3)$, and the value from afterwards (just under four hours later) is $r=1.33(3)$. Error bars for the mean squeezing value are determined using the Student's t distribution. We suspect that these experimental drifts, along with other imperfections in the squeezing operations, are the primary factor limiting HA performance and account for a large portion of our residual phase sensitivity.

We use independent calibration experiments to fix several of the fit parameters, as well as to validate the assumptions we make when fitting. Each calibration experiment consists of preparing a desired initial state of motion, followed by a BSB analysis pulse of varying duration to extract $P_{\downarrow}(t)$. The first calibration experiment performs BSB analysis immediately after the ground state cooling sequence, with no squeezing or displacement applied, as shown in Extended Data Fig.~\ref{fig:calibrations}a. We fit the resulting trace to equation~(\ref{eq:BSB_flopping}) to determine calibrated values of $\gamma$ and $\Omega$, as well as to estimate $\bar{n}$ for the initial motional state after ground state cooling. 

Next we calibrated the squeezing and displacement operations using the fitted values for $\gamma$ and $\Omega$. The squeezing calibration consists of applying the parametric drive to squeeze the initial motional state after ground state cooling, then performing BSB analysis to extract the value of $r$ by fitting to a model for a squeezed ground state, as shown in Extended Data Fig.~\ref{fig:calibrations}c. The duration of the parametric drive is the same as used to generation the squeezing operations in the HA experiments. We calibrate the displacement $\alpha$ by applying a series of resonant motional excitation pulses after first performing ground state cooling, as shown in Extended Data Fig.~\ref{fig:calibrations}b. The number and duration of displacement pulses match the values used in the HA experiments. We fit the measured BSB oscillations to a model for a coherently-displaced ground state. Finally, we characterize the phase coherence of the squeezed states by performing squeezing and then anti-squeezing (the latter by adjusting the phase of the parametric drive) on the ground-state-cooled motional mode and fitting the BSB oscillations to determine the amount of residual squeezing, as shown in Extended Data Fig.~\ref{fig:calibrations}d. Based on the results of our ground state cooling and squeezing/anti-squeezing calibrations, we make the approximation for fitting purposes that the final state after an HA sequence is a coherent state of motion. 

Although for amplification of displacements, the interaction to be amplified does not need to be finely Trotterized (i.e. $N$ does not need to be $>1$) since it commutes with the squeezing operations, we performed the displacement amplification experiment with $N=3$, the results of which are shown in Extended Data Fig.~\ref{fig:amp-disp-n3}. For $N=3$, we split our displacement into 6 pieces rather than 2, which triples the number of squeeze-unsqueeze pairs. This resulted in markedly better phase insensitivity but reduced gain. Increasing the number of squeezing operations increases motional decoherence, but alternating the axes along which we amplify more frequently reduces residual phase sensitivity from imperfections in our squeeze drive.
 
\end{document}